\documentstyle[12pt,epsfig,amssymb]{article}
 \newcommand{\mycaption}[1]{\begin{center}
                            \parbox{4.5in}{
                            \caption{\protect {\footnotesize #1}}}
                            \end{center}
                           }

\def\be{\begin{equation}}
\def\ee{\end{equation}}

\def\l{\left(}
\def\r{\right)}
\def\la{\langle }
\def\ra{\rangle }

\newcommand{\eq}[1]{(\ref{#1})}
\begin{document}
\renewcommand{\thefootnote}{\fnsymbol{footnote}}

%%%%%%%%%%% Titlepage

\begin{titlepage}
\begin{flushright}
\begin{tabular}{l}
ULB--TH--99-31
hep-ph/9912204
\end{tabular}
\end{flushright}
\vskip0.5cm
\begin{center}
  {\Large \bf 
Neutrino masses from nonstandard supersymmetry breaking terms.
  \\}

\vspace{1cm}
{\sc J.-M.\ Fr\`{e}re}${}^{1,}$\footnote{E-mail: frere@ulb.ac.be}, 
{\sc M.V.\ Libanov}${}^{2,1,}$\footnote{E-mail: ml@ms2.inr.ac.ru}, and
{\sc S.V.\ Troitsky}${}^{2,1,}$\footnote{E-mail: st@ms2.inr.ac.ru}
\\[0.5cm]
\vspace*{0.1cm} ${}^1${\it Service de Physique Th\'{e}orique, CP 225,
  Universit\'{e} Libre de Bruxelles, B--1050, Brussels, Belgium}
\\[0.3cm]
\vspace*{0.1cm} ${}^2${\it 
Institute for Nuclear Research of the Russian Academy of
Sciences,
         60th October Anniversary Prospect 7a, Moscow, 117312, Russia
}
\\[1.3cm]

%{\em Version of \today}

  \vfill

  {\large\bf Abstract\\[10pt]} \parbox[t]{\textwidth}{ 
Naturally small Majorana neutrino masses arise from
nonstandard supersymmetry breaking terms. This mechanism
works in the minimal supersymmetric framework and does not require
extra particles or new mass scales.
It could also be responsible for proton decay even in the absence of 
Grand Unification.
}
  \vskip1cm 
\end{center}
\end{titlepage}

\setcounter{footnote}{0}
\renewcommand{\thefootnote}{\arabic{footnote}}

{\bf 1.} In realistic models of broken supersymmetry, two scales usually
appear. One is the scale of SUSY breaking in the hidden sector which
is parameterized by the vacuum expectation value (vev) $F$ of the
auxiliary component of some hidden sector field, another is the scale
$M$ at which SUSY breaking is transferred to the visible sector.

In the gravity mediated
scenario, $M\sim 10^{18}$~GeV. 
Various supersymmetry breaking terms appear in the low energy
lagrangian after integrating out the hidden sector. Soft supersymmetry
breaking terms of MSSM (masses of scalar fields and
trilinear scalar couplings) are of order $F/M$, and 
thus $F/M\sim 1$~TeV to explain gauge hierarchy by
radiative electroweak symmetry breaking.  However, this is not the
whole story and other renormalizable gauge invariant terms could be generated in the
low energy lagrangian. 
The case of dimension $m^1$ terms (e.g., non-holomorphic trilinears)
is well-known. They could be ``hard'' if global singlets are present
(these terms were listed in
the original work on MSSM, ref.~\cite{MSSM}, and were also discussed in
\cite{would-be}).  As was emphasized recently in
ref.~\cite{Martin}, dimensionless couplings may be generated
too. These are hard supersymmetry breaking terms, and such couplings
do induce quadratic divergencies in scalar masses.  This is not
dangerous, however, because all these terms are suppressed by
$F/M^2\sim 10^{-15}$ or even by $F^2/M^4$ (would-be-hard dimensionful
terms are suppressed by $F^2/M^3$). Quadratic divergencies do not
destroy the hierarchy because
corrections to mass scales are highly suppressed and the effective
Lagrangian approach can only be seen with an implicit cutoff. 
Phenomenological relevance of such tiny couplings is doubtful, and
they are usually ignored. 
In ref.~\cite{Martin},
these terms were exploited to stabilize (otherwise) flat directions.
Here, we note that such terms are relevant for the generation of
Majorana neutrino masses of order 
\be	
F^2/M^3\sim 10^{-3}~\mbox{eV}\;.
\label{f2m3}
\ee
Since a Majorana neutrino mass is not invariant under $SU(2)\times
U(1)$, it can be generated only with broken electroweak symmetry, and
thus this term cannot appear at the scale $M$. However, if highly
suppressed couplings violate lepton number, Majorana masses could be
generated radiatively at the electroweak symmetry breaking scale
(at the possible cost of some extra suppression). While the value of
neutrino mass is currently unknown ($m_{\nu_e}<$ a few eV), mass {\em
differences} as low as $\delta m^2=10^{-10}~{\rm eV}^2$ are expected
for the vacuum oscillation solution of the solar neutrino problem.

{\bf 2.} We consider the Minimal Supersymmetric Standard Model (MSSM) with
usual matter content, that is not necessary including right handed
neutrinos. Since the MSSM lagrangian does not contain direct quark--lepton
interactions, it is safe to consider only the electroweak sector (effects
of colored fields are subleading). We allow for the presence of all
renormalizable $SU(3)\times SU(2)\times U(1)$ gauge invariant terms
which are (at least naively) suppressed by $F^2/M^3$ or $F/M^2$, or
weaker. Naive suppression factors can be read out from
ref.~\cite{Martin}. 

With the requirement of $R$ parity conservation, only three
possible couplings are left, with field content $H_U^*\tilde L\tilde E$, $(H_U
H_D)^2$, and $(\tilde L H_U)^2$, where $H_U$ and $H_D$ are Higgs
fields, $\tilde L$ is the left-handed slepton doublet, and $\tilde E$
is the superpartner of $e_L^+$ (or $e_R^-$). 
Once $R$ parity is imposed, only dimensionless couplings can violate
lepton or baryon number. The first two terms  conserve lepton
number, so they cannot generate Majorana mass. The last one, however,
can be used for this purpose. It can be generated in the low energy
lagrangian after integrating out the supersymmetry breaking sector,
for example, from the operator ${1\over M^2}\l XLH_ULH_U\r|_F$, where
$(~)|_F$ denotes an $F$-term. When the auxiliary component of the $X$
superfield developes a vev $(X)|_F\sim F$, this generates the desired
coupling, see ref.~\cite{Martin}.  

Consider the effect of the $SU(2)$ invariant term 
\be
h(\tilde L_i  H_{Uj} \epsilon_{ij})^2,
\label{term}
\ee
where $h$ is the coupling constant of order $F/M^2$ and
$\epsilon_{ij}$ is the usual antisymmetric tensor, $i$ and $j$
are $SU(2)$ indices. Slepton doublets
$\tilde L$ have the same gauge quantum numbers as $H_D$, so this
coupling is easily seen to be gauge invariant. It is also $R$-even,
but breaks lepton number (by a small amount). 

The important observation for us is that, when associated to
electroweak symmetry breaking and non-zero gaugino masses (from soft
supersymmetry breaking), this new term is responsible for the
appearance of Majorana masses for neutrinos. The diagram of
Fig.~1
\begin{figure}
\begin{picture}(0,100)%
\centerline{\epsfxsize=0.4\textwidth \epsfbox{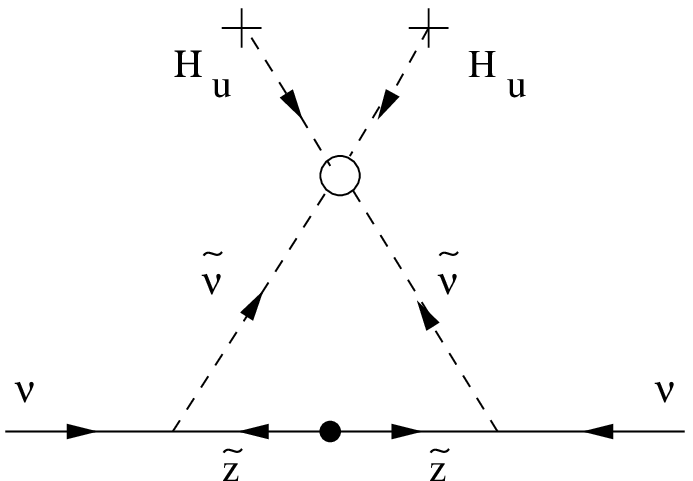}}
\end{picture}%
\mycaption{Contribution to the Majorana neutrino mass. $\tilde Z$ is
zino, crosses at the ends of Higgs lines denote vev, point in the zino
propagator denotes nonzero mass insertion.}
\end{figure}
can be readily evaluated to yield
\be
{h\over 32\pi^2}{g^2 \la H_U \ra ^2\over m_{\tilde\nu} \cos^2\theta_w}
 \,f\!\left(
 {m_{\tilde Z}^2\over
m_{\tilde\nu}^2}  \right),
\label{dia}
\ee
where $h$ is the small coupling constant defined in \eq{term},
$m_{\tilde Z}$ and $m_{\tilde\nu}$ are masses of zino and sneutrino,
respectively, $g$ is the $SU(2)$ coupling constant, $\theta_w$ is the
electroweak angle, and 
$$ 
f(x)={\sqrt{x} (x-1-\log x) \over (x-1)^2},
$$ 
$0.6>f(x)>0.2$ for $0.01<x<100$. The diagram was evaluated in the
case of diagonal $\tilde Z$. We expect this evaluation to be
representative even in the case where $\tilde Z$ mixes with the other
neutralinos (normally decoupled from neutrino).  As Higgs vev and
sneutrino mass are both of order $F/M$, eq.~\eq{dia} indeed gives
neutrino masses as estimated in eq.~\eq{f2m3}.  As expected, the loop
integration implies an extra suppression, here by a factor $16\pi^2$
included in \eq{dia}, which somewhat reduces the result.

The actual value of the mass depends crucially on the unknown hard
coupling $h$ which cannot be determined unless a specific calculable
mechanism of supersymmetry breaking is chosen. 
   
It must be stressed that this contribution appears only as the result
of electroweak symmetry breaking; Majorana mass for the electron is
not gauge invariant and is not generated.

We must now study possible divergent contributions to neutrino masses.
They could appear in higher orders in perturbation theory and require
explicit counterterms for Majorana neutrino masses.  This would signal
that the physical masses are sensitive to unknown dynamics at high
energies, so bare mass terms should be regarded as free parameters of
the theory instead of predictions (a similar problem occurs for MSSM
gauginos, see Ref.~\cite{Frere}).  This is fortunately not the case
for neutrino masses generated in the way discussed above. 

Indeed, we now show that all diagrams contributing to Majorana mass
have negative superficial degree of divergence.  As already noted,
$SU(2)$ breaking is required and thus the contribution must be
proportional to $v$. Because the Majorana mass term has weak
hypercharge 2, and the corresponding diagram has to be gauge invariant
before breaking, at least one extra factor of $v$ is required. (In
practice, we also need to reverse the fermionic flow, which involves
either a gaugino mass or some other dimensionful chirality breaking).

Possible subdivergencies are removed by renormalization of other
parameters of the lagrangian.  Note that the (innocuous, as this
merely modify the new coupling $h$) renormalization of $h$ implied by Fig.~2
\begin{figure}
\begin{picture}(0,100)%
\centerline{\epsfxsize=0.4\textwidth \epsfbox{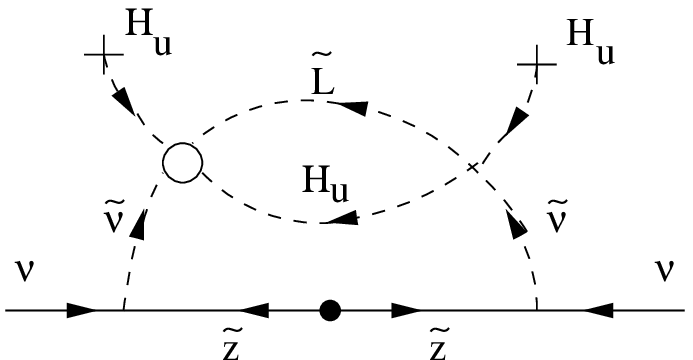}}
\end{picture}%
\mycaption{ Internal loop requiring a renormalization of $h$ in the
presence of Dirac neutrino mass term.  }
\end{figure}
only appears if an
explicit Dirac mass term for the neutrino exists (which requires also 
$\nu_R$ to be included from the start).

% Conclusion

For massless neutrinos, the terms considered here could thus account for (very small) masses
of neutrinos, reminiscent of the vacuum oscillation solution of
the solar neutrino problem. If other contributions to the neutrino
mass exist, these terms could generate a splitting, $\delta m_i\lesssim
10^{-5}$~eV, the resulting $\delta m_i^2$ becomes of order $m_i\delta
m_i$. For a ``common'' neutrino mass around 1~eV (a welcome
contribution to the dark mass of the Universe) this effect could
contribute also to other cases of oscillation.

To summarize, the mechanism which generates the hierarchy $M_{\rm
EW}/M_{\rm Pl}$ can as well generate the hierarchy $m_\nu/M_{\rm
EW}\sim M_{\rm EW}/M_{\rm Pl}$. We present here an explicit
realization of this phenomena in the Minimal Supersymmetric Standard
Model {\it without} additional fields (even right handed neutrinos)
or mass scales. 
Of course, the idea to relate the two hierarchies can work in 
different frameworks. For instance, neutrino masses of order $M_{\rm
EW} \cdot {M_{\rm EW}\over M_{\rm Pl}}$ can be generated 
by extra dimensions with localized gravity, see estimate of
Ref.~\cite{Dudas}. This is however completely different from the
present approach, which relies on supersymmetry breaking and gaugino
masses. 

Our approach results in reasonable neutrino mass
values which are evocative of the vacuum oscillations explanation of
solar neutrino anomaly. It is worth pointing out that the interaction
\eq{term} is flavour-dependent, so the coupling $h$ is
in fact a matrix $h_{ij}$ in the flavour space. The hierarchy of
neutrino masses and mixings in our scenario is completely defined by
this matrix and by the sneutrino masses, and is thus not directly related 
to the mass hierarchy of charged leptons (cf.\ ref.~\cite{Murayama}).

{\bf 3.} 
We now consider baryon number violation and other
dimensionless
supersymmetry breaking terms. Among such couplings are two $R$-even
terms which violate baryon number, namely, $(\tilde Q \tilde Q \tilde
Q \tilde L)$ and $(\tilde U \tilde U \tilde D \tilde E)$ ($SU(3)$
indices are contracted antisymetrically in both terms, $\tilde Q$ is
the squark doublet, and $\tilde U$ and $\tilde D$ are up and down
antisquark singlets). Such terms can of course be excluded from the
onset by
requiring baryon number conservation. It is however
interesting to evaluate their physical impact. These
terms contribute to proton decay through the diagram Fig.~3(a). 
\begin{figure}
\begin{picture}(0,160)%
\centerline{\epsfxsize=0.6\textwidth \epsfbox{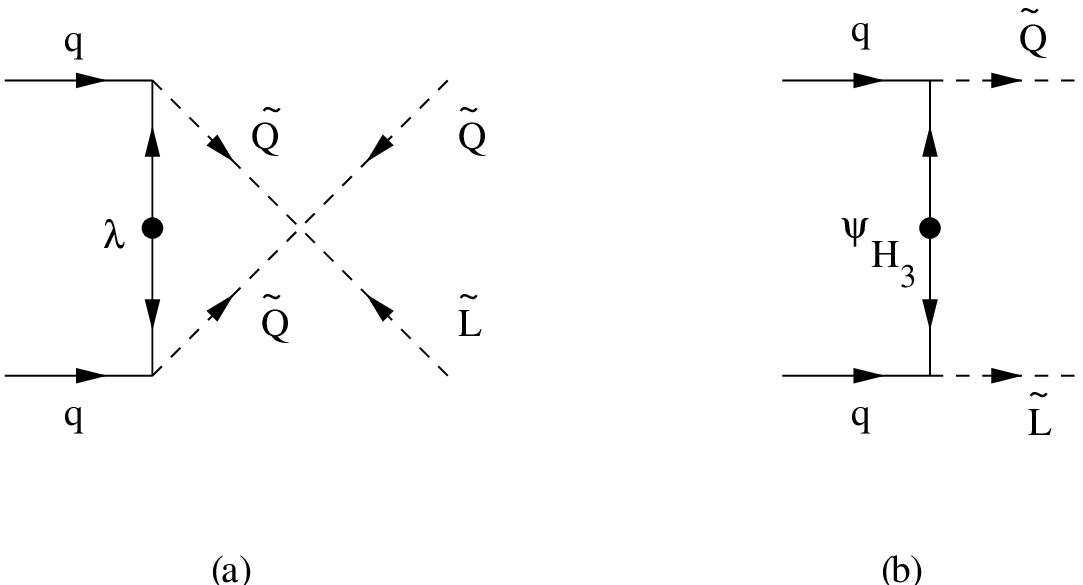}}
\end{picture}%
\mycaption{ 
Operators resulting in proton decay: (a) with a dimensionless
supersymmetry breaking term; (b) with triplet higgsino exchange.
}
\end{figure}
This should be
compared to the usual SUSY GUT contribution from dimension 5 operators
induced by triplet higgsino (the coloured part of the $SU(5)$ 5-plet
Higgs superpartner) exchange, Fig.~3(b). The contribution from 
hard terms, Fig.~3(a), is suppressed by ${F\over M^2}\,{1\over m_{\lambda}}$,
which is numerically of the same order as the GUT contribution,
Fig.~3(b), which is estimated as
$1/m_{\psi_{H_3}} \sim (10^{17}~{\rm GeV})^{-1}$. Note, however, that
proton decay takes place here via hard supesymmetry breaking terms
already in the MSSM, i.e.\ 
without Grand Unification.  

{\bf 4.} In a different context, the effect of nonstandard
supersymmetry breaking terms may also be substantial in models where
supersymmetry breaking is transmitted to the visible sector at low
energies, for instance, in the case of gauge mediation (see
Ref.~\cite{GMSB} for reviews). There, $F$ and $M$ are replaced by the vacuum
expectation values of the auxiliary and scalar components of a chiral
superfield which interacts with messengers, $F/M$ being of order
100~TeV with $M$ much lower than the Planck scale.  This could lead to
larger effects in neutrino masses and rare processes, thus imposing
very strong lower bounds on $M$.

M.L.\ and S.T.\ acknowledge the warm hospitality of the 
Service de Physique Th\'eorique, Universit\'e Libre de
Bruxelles, 
where this work was done. We are indebted to V.A.~Ru\-ba\-kov for
numerous helpful discussions. 
This work is supported in part by the ``Actions de Recherche
Concret\'ees'' of ``Communaut\'e Fran\c{c}aise de Belgique'' and
IISN--Belgium.  
Work of M.L.\ and S.T.\ is supported in part by the
Russian Foundation for Basic Research (RFFI)
grant 99-02-18410a and by the Russian Academy of Sciences, JRP grant
No.~37.

\end{document}